\documentclass[aps,prc,twocolumn,floatfix,superscriptaddress]{revtex4}
\usepackage{epsfig}
\usepackage{amsmath}
\usepackage{color}
\usepackage{marvosym}

\usepackage{graphicx}

\begin{document}

\title{Photon production from Pb+Pb collisions at {$\sqrt{s_{\rm {NN}}}$} = 5.02 TeV at LHC and at {$\sqrt{s_{\rm {NN}}}$} = 39 TeV at FCC}
\author{Pingal Dasgupta}
\email{pingaldg@vecc.gov.in}
\affiliation{Variable Energy Cyclotron Centre, HBNI, 1/AF, Bidhan Nagar, Kolkata-700064, India}
\affiliation{Homi Bhabha National Institute, Training School Complex, Anushakti Nagar, Mumbai 400085, India}
\author{Somnath De}
\email{somvecc@gmail.com}
\affiliation{Pingla Thana Mahavidyalaya, Vidyasagar University, West Bengal-721140, India}
\author{Rupa Chatterjee}
\email{rupa@vecc.gov.in}
\affiliation{Variable Energy Cyclotron Centre, HBNI, 1/AF, Bidhan Nagar, Kolkata-700064, India}
\affiliation{Homi Bhabha National Institute, Training School Complex, Anushakti Nagar, Mumbai 400085, India}
\author{Dinesh K. Srivastava}
\email{dinesh@vecc.gov.in}
\affiliation{Variable Energy Cyclotron Centre, HBNI, 1/AF, Bidhan Nagar, Kolkata-700064, India}
\affiliation{Homi Bhabha National Institute, Training School Complex, Anushakti Nagar, Mumbai 400085, India}
\affiliation{Institut f\"ur Theoretische Physik, Johann Wolfgang Goethe-Universit\"at, Max-von-Laue-Str. 1, D-60438 Frankfurt am Main, Germany}
\affiliation{ExtreMe Matter Institute EMMI, GSI Helmholtzzentrum f\"ur Schwerionenforschung, Planckstrasse 1, 64291 Darmstadt, Germany}

\begin{abstract}
We calculate the production of prompt and thermal photons from Pb+Pb collisions at 5.02A TeV at the Large Hadron Collider (LHC) and at 39A TeV at the proposed  Future Circular Collider (FCC) facility. The photon spectra and anisotropic flow at these energies are compared with the results obtained from 2.76A TeV Pb+Pb collisions at the LHC for three different centrality bins. The prompt photons originating from initial hard scatterings are found to increase by a factor of  1.5 to 2  at 5.02A TeV in the $p_T$ region 2 to 15 GeV and the enhancement is found to be  about 5 to 15 times at FCC energy compared to 2.76A TeV in the same $p_T$ region. The evolution of the Quark-Gluon Plasma (QGP) formed in Pb+Pb collisions at LHC and FCC energies are studied using a hydrodynamical model and the $p_T$ spectra and elliptic flow of thermal photons are calculated using state-of-the-art photon rates. The relative enhancement in the production of thermal photons is found to be more compared to prompt photons at FCC than at the LHC energies. Although the production of direct (prompt+thermal) photons is found to enhance significantly with increase in beam energy, the photon elliptic flow increases only marginally and does not show strong sensitivity to the collision energy.  
\end{abstract}

\maketitle

\section{Introduction} 
Experiments performed at the Relativistic Heavy Ion Collider (RHIC) and at the Large Hadron Collider (LHC) are aimed at exploring a specific region of the  QCD phase diagram where a possible transition from bound state of  hadrons to a unbound state of quarks and gluons  can occur. This color deconfined state of quarks and gluons in local thermal equilibrium is known as Quark-Gluon Plasma~\cite{qgp, sur}.
Relativistic  hydrodynamics has emerged as one of the most successful frameworks to explain the soft probes or bulk observables produced in high energy heavy ion collisions~\cite{uli, hydro}. The charged particle spectra as well as anisotropic flow parameters (elliptic, triangular flow etc.) are successfully explained by hydrodynamic model with suitable initial conditions where the initial conditions are  constrained from the experimental data for final charged particle multiplicity~\cite{fl1, fl2, v23, v3}. 

Photons, both real as well as virtual (i.e., dileptons) are known as one of the promising probes to study the hot and dense Quark Gluon Plasma produced in relativistic heavy ion collisions~\cite{phot1, phot2}. The direct photon spectra at 200A GeV Au+Au collisions at RHIC and at 2.76 A TeV Pb+Pb collisions at the LHC are  explained well in the region $p_T > $ 1 GeV by theory calculation combining the contributions of prompt and thermal photons where the prompt photons are calculated using a next-to-leading order (NLO) perturbative QCD calculation and the thermal part is calculated considering a hydrodynamic evolution of the produced fireball and state of the art photon production rates~\cite{chre3, chre4,osu,phsd}. However, the very low $p_T$ ( $\le$ 1 GeV) region of the direct photon spectra which is likely to be  dominated by photons produced from the interaction of different hadronic channels still remain unexplained by most of the theory calculations. 
Most importantly, it has been shown in many recent studies that the theory calculations underpredict the experimental data on the elliptic and triangular flow of photons by a large margin both for Au+Au collisions at RHIC and Pb+Pb collisions at the LHC~\cite{chre3,chre4,dcs}. Thus, model calculations that simultaneously explain the spectra and anisotropic flow of charged particles from heavy ion collisions, fail to reproduce the photon spectra and anisotropic flow parameters at RHIC~\cite{phenix_phot, phenix_v2} and LHC energies~\cite{alice_phot, alice_v2}. This is known as {\it {photon $v_2$ puzzle}}. The calculation of photon anisotropic flow parameter at higher collision energies, at different collisions centralities as well as for different systems (e.g., Cu+Cu, U+U) with modified initial conditions would be valuable to understand this puzzle~\cite{dcs,uu,shadow}. 

It is to be noted that the initial parameters which play a significant role in hydrodynamic model calculations, the formation time, initial temperature, freeze-out temperature etc. are not known precisely till date. We also know that the photon spectra are much more sensitive to the initial state of the produced fireball than the hadron spectra as the hadrons are only emitted from freeze-out surface, whereas, photons are emitted  through out the system evolution~\cite{cs,chre2}. Thus, we calculate the production and anisotropic flow of direct photons at $\sqrt{s_{\rm{NN}}}$=5.02 TeV,  the highest energy for Pb+Pb collisions achieved at LHC till date.  We expect that Pb+Pb collisions at 5.02A TeV would be quite valuable to constrain the initial conditions  as well as helpful to understand the discrepancy between the theory calculation and experimental data on photon elliptic flow parameter. The inverse slope of the photon spectra at 5.02A TeV would provide the effective temperature of the produced QGP matter at that energy. In this study we estimate the production of prompt photons at 5.02A TeV Pb+Pb collisions for three different centrality bins and compare those with the photon results obtained from 2.76A TeV Pb+Pb collisions at the LHC. In addition, we calculate the production and elliptic flow of thermal photons at 5.02A TeV for different centrality bins.

The proposed Future Circular Collider (FCC) aims to collide protons at 100 TeV. Heavy ion collision at FCC energy is a part of accelerator design study~\cite{fcc, fcc1, fcc2}. The Pb+Pb collisions at FCC are expected to happen at 39A TeV, which is more than 7 times larger than the top LHC energy achieved for Pb+Pb collisions till date. Predictions from hydrodynamic model calculation have shown that the  charged particle multiplicity, life-time and volume of the produced fireball  (for most central Pb+Pb collisions) would increase by a significantly large factor at FCC compared to LHC energies~\cite{fcc2}. One can also expect to see a large enhancement in the production of direct photons at FCC  compared to the LHC. We have seen a marginal enhancement in the photon elliptic flow  at LHC compared to RHIC. Thus, the estimation of photon $v_2$ at FCC would be valuable to conclude about how sensitive  the anisotropic flow parameter is to the beam energy of heavy ion collisions. In addition, the significant enhancement in photon production at FCC is also expected to reduce the large error bars in the experimental  photon $v_2$ data and this  would be helpful in understanding the discrepancy between the experimental data and results from theory calculation.

We predict the production of direct photons (prompt and thermal) from Pb+Pb collisions at 39A TeV at FCC and compare with the results obtained at the two (2.76A TeV and 5.02A TeV) LHC energies. The elliptic flow parameter calculated at FCC would provide an upper limit of the photon $v_2$ which can be achieved in heavy ion collisions. We shall see that the prompt photons and thermal photons as well as the elliptic flow of thermal photons change by differing extents as the energy of the collisions  and the resulting  initial conditions are changed. Thus an accurate description of the direct photon (sum of the thermal and prompt) spectrum along with the elliptic flow can constrain our description for these.

The paper is organized as follows. We discus the production of prompt photons in Section II. The hydrodynamics model and thermal photon calculations are discussed in the Section III. In Section IV we show the results of photon spectra and elliptic flow parameter. The summary and conclusions are given in Section V.
  
\section{Prompt photons}
Prompt photons which are produced from initial hard scatterings of the colliding nucleons are the dominant source of direct photons in the high $p_T$ region ($p_T \ge$ 4 GeV).  Quark-gluon Compton scattering ($q$ ($\bar q$) + g $\longrightarrow$ $q$ ($\bar q$) + $\gamma$), quark-anti-quark annihilation ($q$ + $\bar q$ $\longrightarrow$ g + $\gamma$) and 
bremsstrahlung emission from final state partons ($q$($\bar q$) $\longrightarrow$ $q$($\bar q$) + $\gamma$) are the leading production channels 
of prompt photons. The photons emitted in the first two reactions are known as {\it direct} 
prompt photons and those emitted from bremsstrahlung process are known as  {\it fragmentation} photons.
The prompt photon production cross section in elementary hadron-hadron (A+B) collisions can be expressed as~\cite{arleo}:
\begin{equation}
\begin{aligned}
\frac{d^2\sigma^{\gamma}}{d^2p_{T}dy}=\sum_{i,j}\int dx_{1}f^{i}_{A}(x_{1},Q^2_f)\int dx_{2}f^{j}_{B}(x_{2},Q^2_f) \\
\times \sum_{c=\gamma,q,g}\int \frac{dz}{z^2}\frac{d\sigma_{ij\rightarrow cX}(x_1,x_2;Q^2_R)}{d^2p^c_T dy_c} D_{c/\gamma}(z,Q^2_F),
\end{aligned}
\label{one}
\end{equation} 
where, $f^i_A(x_{1},Q^2_f)$ is the parton distribution function (PDF) of $i^{\rm{th}}$ (flavored)  parton in hadron A carrying a momentum fraction $x_1$. 
Similarly, $f^j_B(x_{2},Q^2_f)$ corresponds to the PDF for  $j^{th}$ (flavored) parton in hadron B carrying a momentum fraction $x_2$. 
$Q_f$ is the factorization scale appearing from the QCD factorization scheme~\cite{Collins,R.D.Field}. $ D_{c/\gamma}(z,Q^2_F)$ denotes the  parton to photon 
vacuum fragmentation probability defined at $z={p_\gamma}/{p_c}$ and $Q_F$ is the fragmentation scale also appearing under the same QCD factorization scheme. 
 The fragmentation function reduces to $\delta(1-z)$ when a photon is emitted in the direct process i.e., $c=\gamma$. The term 
$\sigma_{ij\rightarrow cX}(x_1,x_2;Q^2_R)$ signifies the hard parton-parton cross-section for all the relevant processes in which a photon 
is produced either directly or fragmented off the final state partons (q or g). $Q_R$ is the momentum scale which appears due to the renormalization 
of the running coupling constant $\alpha_s(Q^2)$. \\
For the nucleus-nucleus (A+A) collisions  we replace the 
elementary nucleon PDF (see Eq.\ref{one}) by the isospin averaged nuclear PDF:
\begin{equation}
\begin{aligned}
f^i_A(x,Q^2)=R_A(x,Q^2)\Big[ \frac{Z}{A}f^i_p(x,Q^2)+\frac{A-Z}{A}f^i_n(x,Q^2)\Big],
\end{aligned}
\end{equation}
where, $R_A(x,Q^2)$ is the nuclear modification to the PDF~\cite{EMC} and $f^i_p$, $f^i_n$ are the free proton and neutron PDFs respectively. 
We have used EPS09 parameterization~\cite{eps09} of nuclear shadowing function in this study.
$Z$ and $A$ are the atomic number and atomic mass respectively of the colliding nucleus. 
For a non-central collision at impact parameter $b$ we replace $Z$ and $A$  by the effective atomic and 
mass number respectively using the relation~\cite{centpr}:
\begin{equation}
\begin{aligned}
Z_{\rm{eff}}=\frac{Z}{A} \frac{N_{\rm{part}}(b)}{2}, \   N_{\rm {eff}}=\frac{N}{A} \frac{N_{\rm {part}}(b)}{2} \, .
\end{aligned}
\end{equation}
Here $N_{\rm {part}} (b)$ is the number of participant (or wounded) nucleons in an A+A collision at impact parameter $b$ calculated using Glauber model formalism.  
The prompt photon invariant yield is obtained from the differential production cross-section in nucleon-nucleon (nn) collisions as:
\begin{equation}
\begin{aligned}
\frac{d^2N^{\gamma}_{\rm AA}}{d^2p_{T}dy}=\frac{d^2\sigma_{nn}^{\gamma}}{d^2p_{T}dy} \times T_{\rm AA}(b),
\end{aligned}
\end{equation}
where, $T_{\rm AA}(b)$ is the nuclear overlap function. \\

In the present study, we estimate the prompt photon production from Pb+Pb collisions at the mid-rapidity  ($|y|<$ 0.5) using 
the CTEQ6.6 parton distribution functions~\cite{cteq} and BFG-II photon fragmentation functions~\cite{bfg}. We have used the 
Monte Carlo code JETPHOX (version 1.2.2)~\cite{Jetphox} which includes all leading order 
 and the next-to-leading order (in $\alpha_s$) 
 channels of prompt photon production~\cite{Owens}.
We  consider $Q_R$, $Q_f$ and $Q_F$ to be same ($= Q$) and all equal to the $p_T$ of photons.  
One can fine tune these scales to reproduce the experimental prompt photon spectra in  p+p collisions.  
It has been shown in~\cite{sdks_rhic} that calculation considering a scale value of $p_T/2$ agrees well with the experimental 
data from 200 GeV p+p collisions at RHIC. However,  at 2.76 TeV at LHC the data matches with the result from theoretical calculation for a scale value of $p_T$ of the produced photons~\cite{sd_phot2}.  In absence of any better guideline for choosing these scales at the higher LHC energies or the FCC energies, we have chosen the same scale, i.e.,  $Q=p_T$ for these energies as well.
\\

\begin{figure}
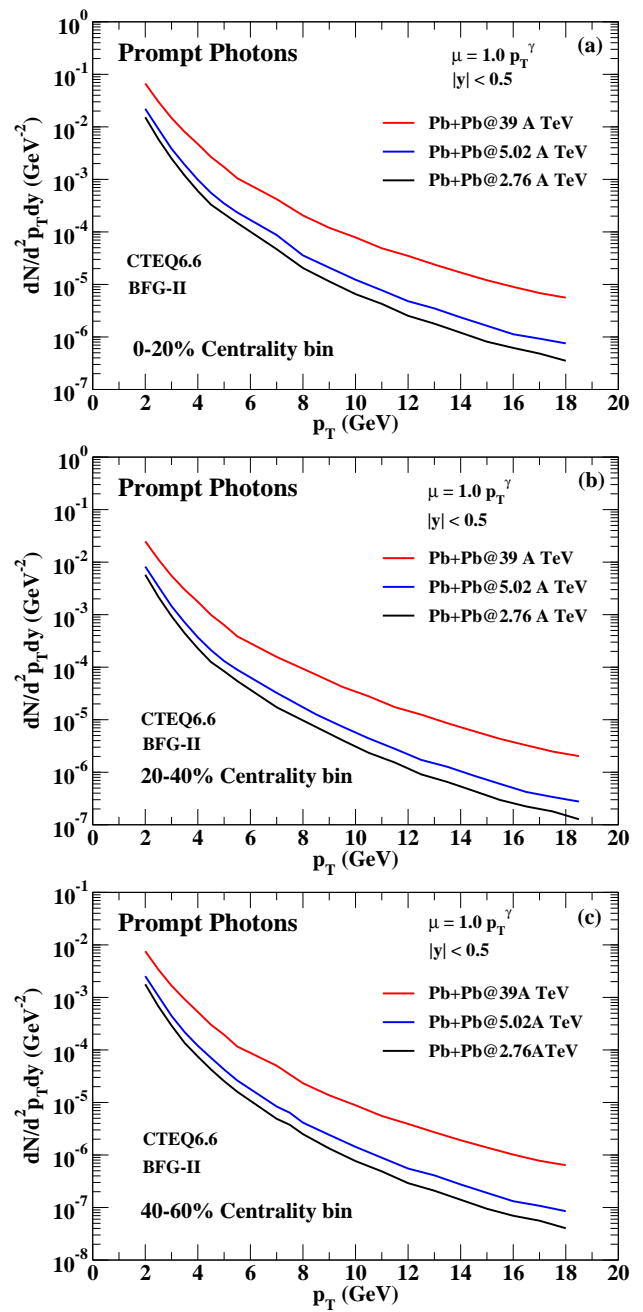

\centerline{\includegraphics*[width=8.2 cm]{0_20_prompt.eps}}
\centerline{\includegraphics*[width=8.2 cm]{20_40_prompt.eps}}
\centerline{\includegraphics*[width=8.2 cm]{40_60_prompt.eps}}
\caption{(Color online) Prompt photon spectra from Pb+Pb collision at 2.76 and 5.02A TeV at LHC and at 39A TeV at FCC  for centrality bins 0--20\% (a), 20--40\% (b), and 40--60\% (c).}
\label{fig1}
\end{figure}
\begin{figure}
\centerline{\includegraphics*[scale=0.85,clip=true]{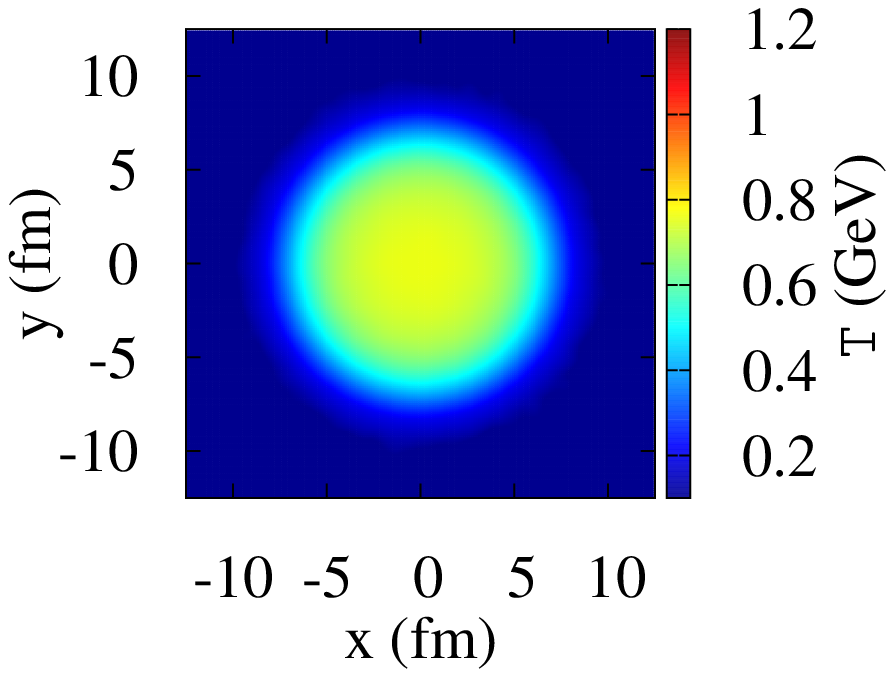}}
\centerline{\includegraphics*[scale=0.85,clip=true]{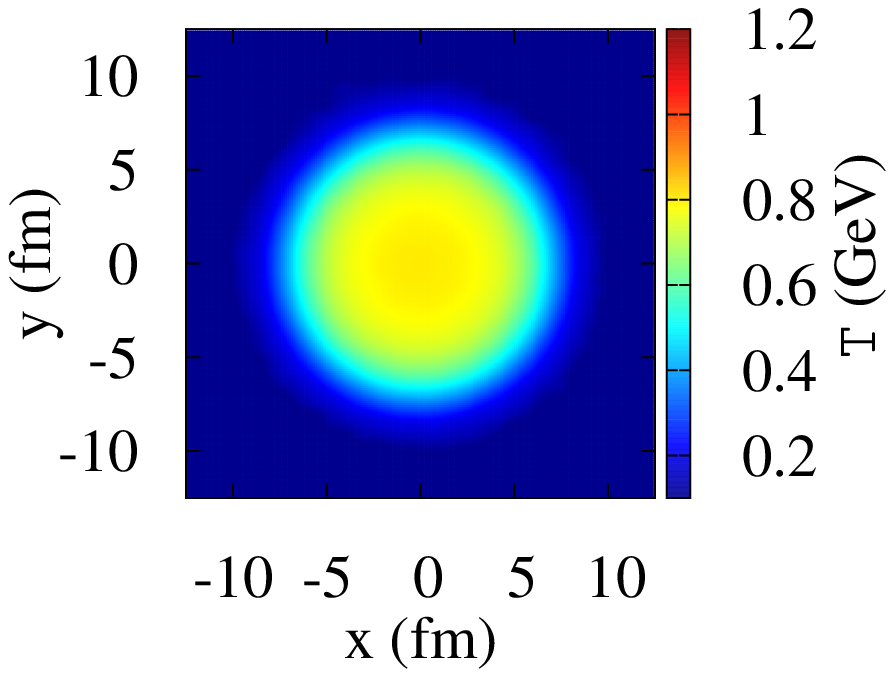}}
\centerline{\includegraphics*[scale=0.85,clip=true]{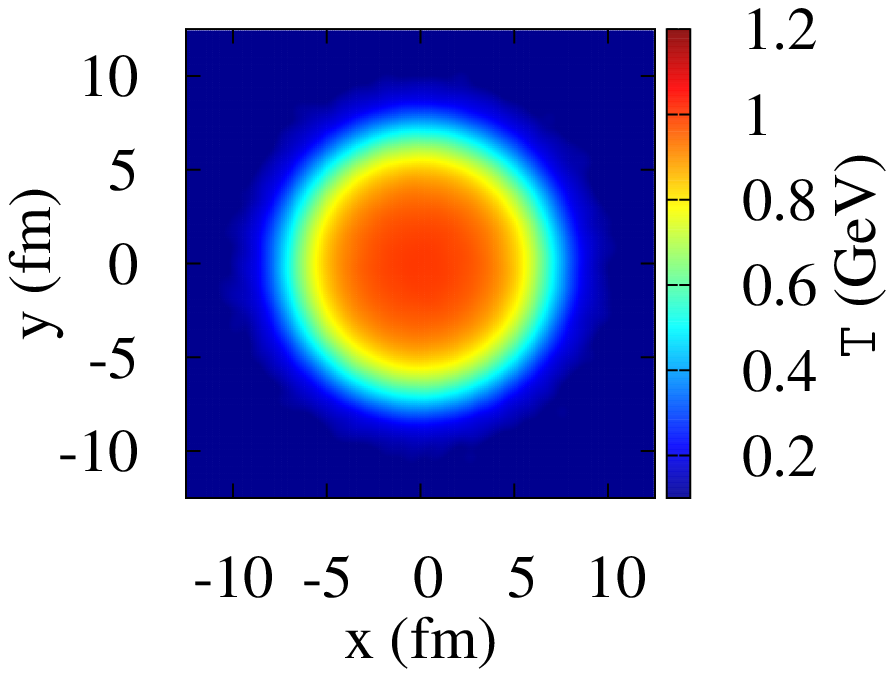}}
\label{fig2.1}
\caption {(Color online) Distribution of temperature at the formation time $\tau_0$ (taken as 0.14 fm/$c$) on transverse ($x-y$) plane for central (b $\approx$ 0 fm ) Pb+Pb collisions at 2.76A TeV (upper panel) and 5.02A TeV (middle panel) at LHC and at 39A TeV at FCC (Lower panel). Color bars are the index of temperature in GeV.}
\end{figure}
\section{Hydrodynamic framework and initial conditions}
We have considered a longitudinally boost invariant  (2+1) dimensional ideal hydrodynamic framework~\citep{hannu} with a smooth initial density distribution 
 to study the evolution of the  hot and dense fireball produced in Pb+Pb collisions at relativistic energies. It has been shown in earlier studies that the effect of fluctuations is found to be less pronounced for heavy ion collisions at 2.76A TeV  than at 200A GeV~\cite{chre2}. In addition, event-by event fluctuating initial conditions are found to affect  the anisotropic flow of photons  significantly more for Cu+Cu collisions compared to Au+Au collisions at RHIC~\cite{dcs}. As a result, the effect of initial state fluctuation on the thermal photon production and its anisotropic flow is expected to be less significant at 5.02A TeV and at 39A TeV  than at 2.76A TeV  Pb+Pb collisions at the LHC. However, we would like to mention that a calculation considering the event-by-event fluctuating initial conditions  would be valuable  to get a precise estimation of the production of direct photons at these very high energies and we postpone that for a future study. In our calculation, an entropy initialized smooth transverse profile is constructed by taking average over a sufficiently large number of Monte Carlo (MC) events as follows\citep{shadow} :
\begin{equation}
\begin{aligned}
\label{eq.ncollex}
s(x,y) = \frac{1}{N} \sum_{j=1}^{N} s_{j}(x,y) \, .
\end{aligned}
\end{equation} 
where, $s_{j}(x,y)$ denotes the transverse entropy density  profile for a single MC event. For simplicity we consider an wounded nucleon (WN) profile and the initial entropy density is distributed in the transverse plane  using the relation,
\begin{equation}
\begin{aligned}
\label{imple_eqn}
s_{j}(x,y)=K\sum _{i=1}^{N_{wn}} f_i(x,y)
\end{aligned}
\end{equation}
 $N_{\rm{wn}}$ is the total number of wounded nucleons in an event. $K$ is a constant factor that is tuned from the final charged particle multiplicity and $f_i(x,y)$ is a two dimensional Gaussian distribution function of the form:
 \begin{equation}
  f_i(x,y) = \frac{1}{2 \pi \sigma^2} \exp \Big( -\frac{(x-x_i)^2+(y-y_i)^2}{2 \sigma^2} \Big)
 \label{eq:eps}
\end{equation}
where, $(x_i,y_i)$ is the position of the $i^{\rm{th}}$ source in the transverse plane.  The parameter $\sigma$ decides the size of initial density fluctuation and we use $\sigma = 0.4$ fm for our calculations \cite{hannu, chre1}. 

We take the inelastic nucleon-nucleon cross-section ($\sigma_{\rm {NN}}$) as 42 mb, 64 mb, and 80 mb at the beam energies 2.76A TeV, 5.02A TeV, and 39A TeV respectively~\cite{chre4,fcc2}. The final charged particle multiplicity ($dN_{\rm {ch}}/d\eta$) for 0--5\% centrality class from which the overall normalization constant (K in Eq.\ref{imple_eqn}) is fixed, is taken as 2000 at 5.20A TeV and 1600 at 2.76A TeV Pb+Pb collisions at LHC from the experimental data. For FCC,  the final charged particle multiplicity for most central collisions is estimated by extrapolating the results  at RHIC and LHC energies using the relation $dN_{\rm{ch}}/d{\eta} \propto (\sqrt{s_{\rm{NN}}})^{0.3}$. We get ($dN_{\rm {ch}}/d\eta$) as 3600 at FCC which can be considered as an upper limit of the charged particle multiplicity value~\cite{fcc2}.

The initial thermalization time of the hydrodynamic evolution is taken as $\tau_0$ = 0.14 fm/c for 2.76A TeV collisions~\cite{chre4} and  we retain the same value for $\tau_0$ even for the higher beam energies considered in the present study. One may argue that the system would take smaller time to thermalize for higher beam energies. However, we believe that this value of $\tau_0$ is already too small and in addition at present we do not have any result from theoretical calculation predicting the formation time at 5.02A TeV and at 39A TeV. Thus,  $\tau_0$ = 0.14 fm/c can be considered as a good approximation of $\tau_0$ for all three energies.
The temperature at freeze-out ($T_{f}$) is taken as 160 MeV which reproduces the measured $p_T$  spectra of charged pions at 2.76A TeV at LHC energy. The value of quark-hadron transition temperature ($T_{c}$) is taken as 170 MeV and the lattice QCD based EoS is taken from~\cite{eos}. We choose centrality cuts in our calculation  using the MC Glauber model.

It is important to mention here  that a  two-component (combination of wounded nucleons and binary collisions) Glauber model initial condition is more effective (than an wounded nucleon profile) for  very high energy A+A collisions where we fix the fraction of the two components from the charged particle multiplicity distribution. However, at FCC energy as there is no experimental data available, we consider  single component Glauber model to estimate the thermal photon production at different centralities to avoid introduction of an additional parameter. 
\section{THERMAL PHOTONS} 
We use complete leading order plasma rates from~\cite{amy} to calculate the photon production from the QGP phase.  
The rates for photon production from hadronic phase (an exhaustive set of hadronic reactions and radiative
decay of higher resonance states are considered) have been
taken from ~\cite{trg} and which also include the effects of
the hadronic form factors.
The $p_T$ spectrum of thermal photons is obtained by integrating the emission rates ($R=EdN/d^3pd^4x$) over the entire space-time  history. The evolution is considered from the initial 
thermalization time to the final freeze-out state of the fireball via intermediary quark-hadron 
transition:
\begin{equation}
E \frac{dN}{d^3p}= \int d^4x \, R \left(E^*(x),T(x)\right).
\label{emrate}
\end{equation}
\\
where $T(x)$ is the local temperature. The energy in the comoving frame is $E^* (x)$ = $p^\mu u_\mu (x)$ where, $p^\mu$ is the four-momentum of the photons and $u_\mu$ is the  local four-velocity of the flow field. The values of $T$ and $u_\mu$ are obtained by solving the hydrodynamical equations.

The anisotropic flow co-efficients $v_n$ are estimated by expanding the invariant particle distribution in transverse plane using Fourier decomposition:
\begin{equation}\label{eq: v2}
\frac{dN}{d^2p_Tdy} = \frac{1}{2\pi} \frac{dN}{ p_T dp_T dy}[1+ 2\, \sum_{n=1}^{\infty} v_n (p_T) \, \rm{cos} \,  (n\phi)] \, 
\end{equation}
For smooth initial density distribution the first non-vanishing anisotropic flow coefficient is $v_2$ or the elliptic flow parameter.
\begin{figure}
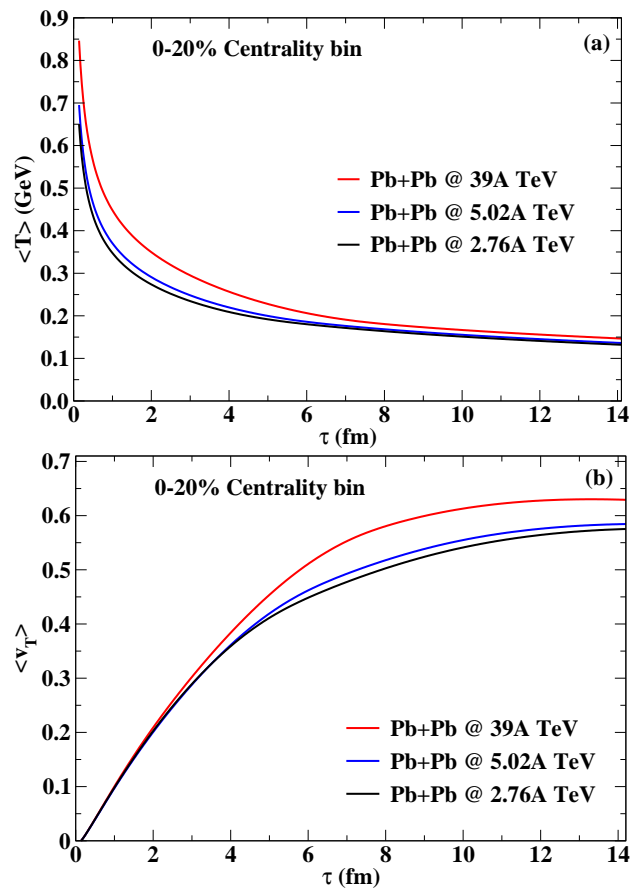

\centerline{\includegraphics*[width=8.2 cm]{avg_temp.eps}}
\centerline{\includegraphics*[width=8.2 cm]{avg_vt.eps}}
\label{fig3}
\caption{(Color online) Time evolution of average temperature (a) and average transverse flow  velocity (b) from 0--20\% central collisions of Pb+Pb  at 39A TeV, 5.02A TeV and 2.76A TeV.}
\end{figure}

\begin{figure}
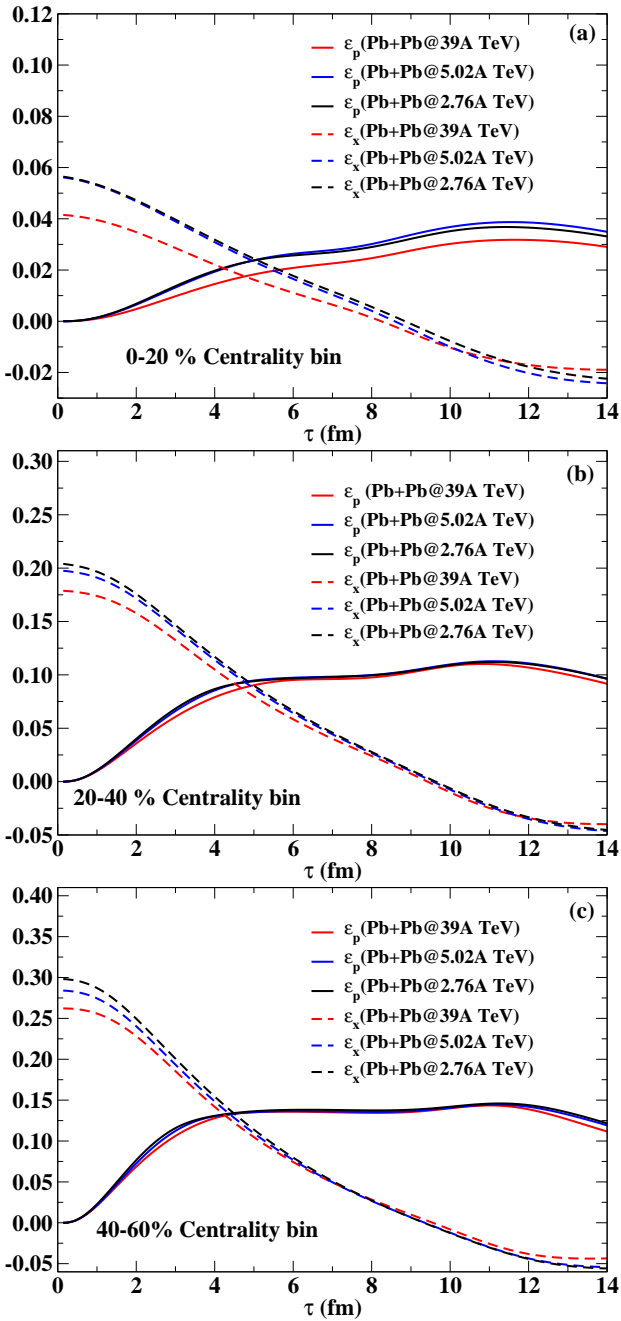

\centerline{\includegraphics*[width=8.2 cm]{0_20_epex.eps}}
\centerline{\includegraphics*[width=8.2 cm]{20_40_epex.eps}}
\centerline{\includegraphics*[width=8.2 cm]{40_60_epex.eps}}
\caption{(Color online) Time evolution of spatial and momentum anisotropies in Pb+Pb collisions at 2.76A TeV and 5.02A TeV  at LHC and at 39 A TeV at FCC at for centrality bins 0--20\% (a), 20--40\% (b), and 40--60\% (c).}
\label{fig4.6}
\end{figure}

\begin{figure}
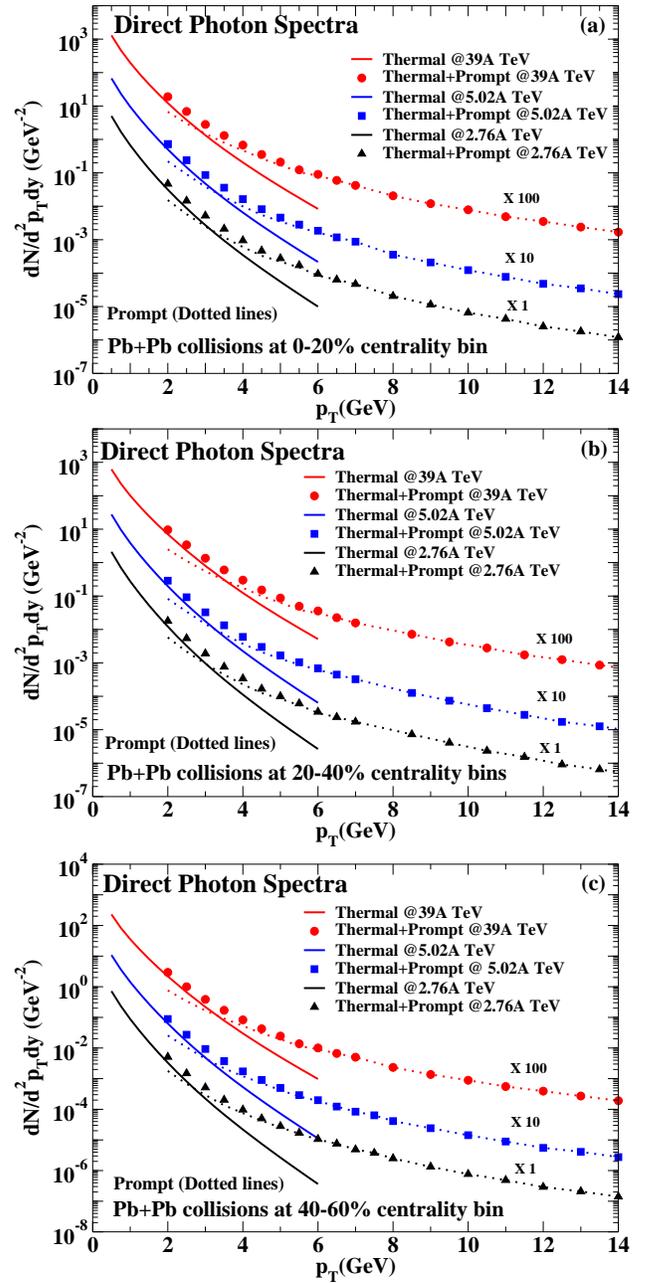

\centerline{\includegraphics*[width=8.2 cm]{0_20_direct.eps}}
\centerline{\includegraphics*[width=8.2 cm]{20_40_direct.eps}}
\centerline{\includegraphics*[width=8.2 cm]{40_60_direct.eps}}
\caption{(Color online) Thermal and thermal+prompt photon spectra from Pb+Pb collisions at 2.76A TeV and 5.02A TeV  at LHC and FCC at 39A TeV for centrality bins 0--20\% (a), 20--40\% (b), and 40--60\% (c).}
\label{fig4}
\end{figure}

\section{Results}
\subsection{Prompt photon production}
The prompt photon spectra from Pb+Pb collisions at 5.02A TeV at LHC and 39A TeV at FCC  for centrality bins 0--20\%, 20--40\%, and 40--60\% are shown in Fig.\ref{fig1}. 
The results from 2.76A TeV are also shown in the same figures for a comparison. For 0--20\% centrality bin one can see that in the  $p_T$ range 2--15 GeV the production of prompt photons is about 1.5--2 times larger at 5.02A TeV compared to 2.76A TeV. 
We see a much larger production of prompt photons at the FCC compared to the LHC,  at $p_T \sim$ 2 GeV the  spectra (for all three centrality bins) at FCC is about 5 times larger than at 2.76A TeV. 
As we move towards  higher $p_T$ values, the enhancement in the production at FCC compared to LHC is even more. At $p_T \sim$ 15 GeV, the enhancement factor is about 15 for prompt photons at 39A TeV than at 2.76A TeV. However,  the difference between spectra at the two LHC energies remains almost same in the entire $p_T$ range shown in the figure. Prompt photon yield for 0--20\% central collisions is found to be almost 9 to 10 times larger than the same obtained for 40--60\% centrality  and almost 3 to 4 times larger for 20--40\% collision centralities for all beam energies. However, the relative enhancement in the production at the three collision energies is found to be similar for all three centrality bins. 
 These variations are well beyond that due to the variation in number of collisions ($N_{\rm{coll}}$).
  
\begin{figure}
\centerline{\includegraphics*[width=8.2 cm]{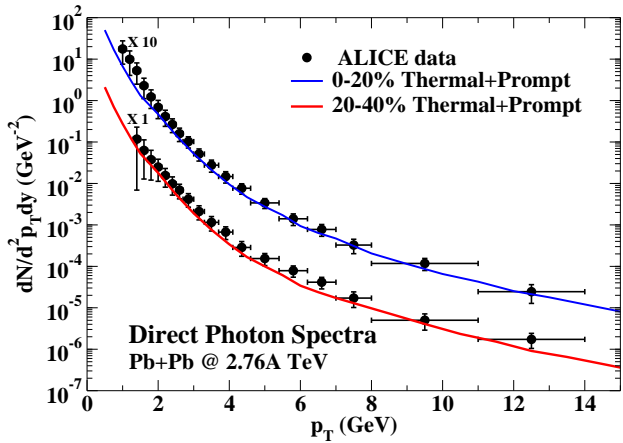}}
\caption{(Color online) Direct photon spectra for 0--20\% and 20--40\% centrality bins at 2.76A TeV Pb+Pb collisions at LHC along with ALICE data~\cite{alice_phot}. }
\label{fig4.5}
\end{figure}
\subsection{Hydrodynamic evolution of the hot and dense matter produced at LHC and FCC energies}
The distribution of initial temperature on the transverse plane for most central (b $\approx$ 0) collision of Pb nuclei at 2.76A TeV, 5.02A TeV and 39A TeV is shown in Fig.2. The value of $\tau_0$ is taken as  0.14 fm/$c$ (as discussed earlier) for all three cases. The smooth initial temperature distribution as shown in the figure is obtained by averaging over 10000 events with fluctuating initial density distributions (using Eq.~\ref{eq.ncollex}). Color bars shown alongside the figure indicate the temperature values. The temperature profile at 5.02A TeV looks hotter than the profile  2.76A TeV as expected while at the FCC energy the central region shows significantly larger temperature compared to both the LHC energies.  At FCC the central temperature (at x=y=0) is found to be more than 1 GeV, which is significantly larger than the maximum central temperatures at the LHC energies.
Hence, more prominent QGP signatures are expected to be obtained from the hotter and longer lived QGP phase at FCC compared to the LHC energies.
\begin{figure}
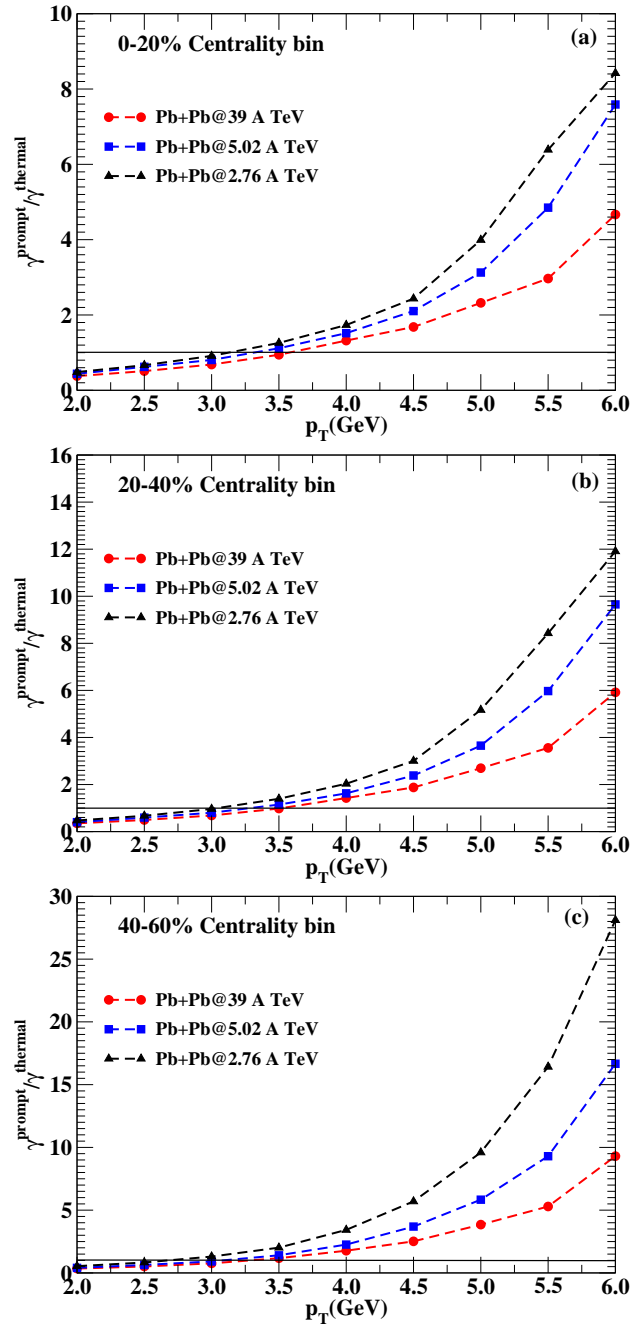

\centerline{\includegraphics*[width=8.2 cm]{pr_th_ratio_0_20.eps}}
\centerline{\includegraphics*[width=8.2 cm]{pr_th_ratio_20_40.eps}}
\centerline{\includegraphics*[width=8.2 cm]{pr_th_ratio_40_60.eps}}
\caption{(Color online) Ratio of prompt and thermal photon production from Pb+Pb collisions at 2.76A TeV and 5.02A TeV  at LHC and FCC at 39A TeV for centrality bins 0--20\% (a), 20--40\% (b), and 40--60\% (c).}
\label{fig6.6}
\end{figure}
\begin{figure}
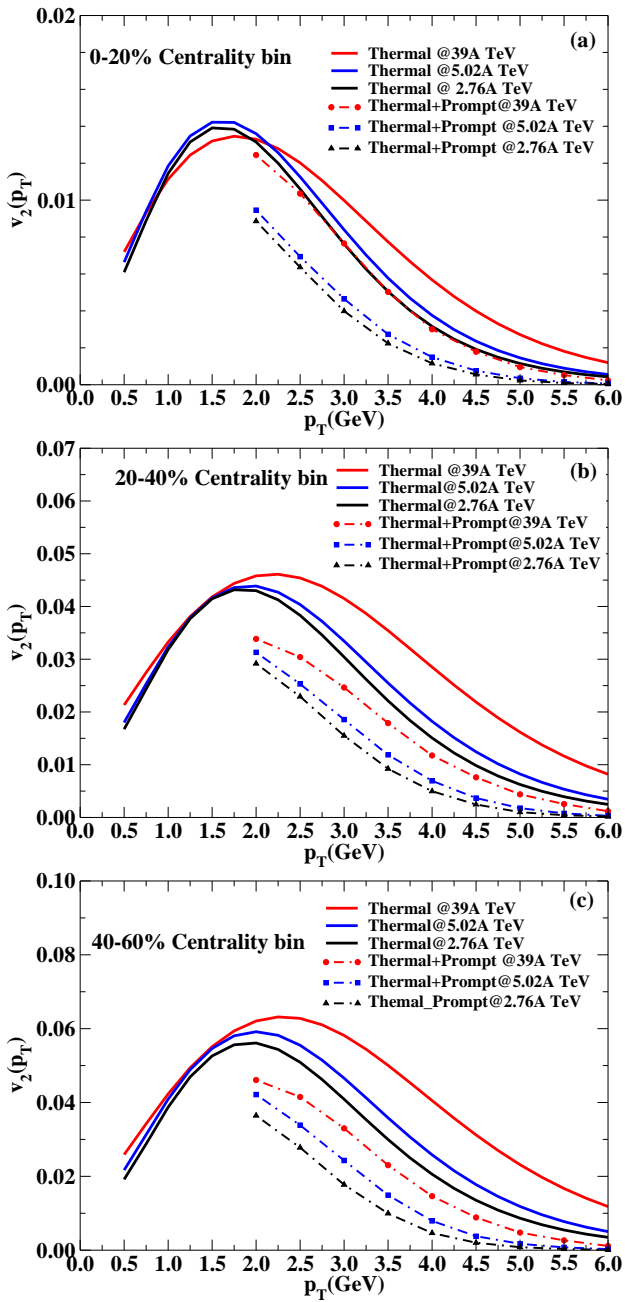

\centerline{\includegraphics*[width=8.2 cm]{v2_0_20_dir.eps}}
\centerline{\includegraphics*[width=8.2 cm]{v2_20_40_dir.eps}}
\centerline{\includegraphics*[width=8.2 cm]{v2_40_60_dir.eps}}
\caption{(Color online) Thermal and direct (thermal+prompt) photon $v_2$ from Pb+Pb collisions at 2.76A TeV and 5.02A TeV at LHC and at 39A TeV at FCC for centrality bins 0--20\% (a), 20--40\% (b), and 40--60\% (c).}
\label{fig5}
\end{figure}
The time evolution of average temperature  and average transverse flow velocity for the three energies (at centrality bin  0--20\%) are shown in Fig. 3. The averages of the thermodynamic quantities are calculated using Eq. (8) of Ref.~\cite{shadow}. The $\langle T \rangle$ at the FCC is about 850 MeV whereas, it is about 690 MeV and 650 MeV  at 5.02A TeV and 2.76A TeV respectively at time $\tau_0$. We see that the average temperature is significantly larger at FCC compared to the LHC energies throughout the evolution of the fireball. A sharp fall in $\langle T \rangle$ is observed for $\tau <$ 4 fm/$c$ for all three energies.  However, the much larger value of $ \langle T \rangle$ even after 4 fm time period at FCC implies that the  QGP phase is longer lived at FCC than at LHC. 
The rise in $\langle v_T\rangle$ with $\tau$ is found to be similar in the first 2--3 fm/$c$ time period for FCC and LHC  energies. However, for $\tau >$ 4 fm/$c$ the average transverse flow velocity rises at a faster rate at FCC than at the LHC energies. 

We show the time evolution of the spatial ($\epsilon_x$) and momentum ($\epsilon_p$) anisotropies (calculated using Eq. (6) and (7) respectively of Ref.~\cite{chre3}) at the three collision energies and  for all three different centrality bins in Fig.~\ref{fig4.6}. The spatial anisotropy is found to be little large for the lowest beam energy at all three centrality bins. On the other hand, the $\epsilon_p$ is found to be similar for all three centrality bins except for the 0--20\% central collisions where it is smallest for FCC energy. 
\subsection{Direct photon production}
The thermal photon spectra from Pb+Pb collisions at the FCC and LHC energies are shown in Fig.~\ref{fig4} for centrality bins 0--20\%, 20--40\%, and 40--60\%. The direct photon spectra obtained by adding the prompt and thermal photons together are also shown in the same figures for a comparison. 
The photon spectra from our calculation are found to explain the ALICE direct photon data~\cite{alice_phot} for Pb+Pb collisions at 2.76A TeV well for 0--20\% and  20--40\% centrality bins as shown in Fig.~\ref{fig4.5}.
The thermal radiation which dominates the direct photon spectrum upto 3--4 GeV  increases  significantly  at FCC compared to the LHC energies.  At $p_T \sim $ 1 GeV, the photon spectra  at 5.02A TeV are found to be almost 1.5 times larger compared to the results at 2.76A TeV whereas, at FCC energy those are  almost 2.8 times larger than at 2.76A TeV at the same $p_T$ value. 

 In order to understand the relative contributions of the prompt and thermal photons at different beam energies and $p_T$ bins, we plot their ratio as a 
function of $p_T$ in Fig.~\ref{fig6.6}. One can see that the ratio is $<$ 1 at $p_T \sim$ 2.0 GeV and  
then rises for higher $p_T$ values. The radiations from thermal medium and initial hard scatterings become equal at $p_T \sim$ 3.5 GeV
at FCC and around $p_T \sim $ 3.0 GeV at both LHC energies for most central Pb+Pb collisions. 
The prompt contribution starts dominating over the thermal radiation as we move towards  higher $p_T$ values and the value of the ratio increases much more rapidly. For $p_T >$ 5 GeV the prompt photons completely outshine the thermal radiation and  at this $p_T$ the ratio is much larger than 1  for all the three energies.
Thus, we conclude that the relative enhancement in thermal photon production compared to the prompt photons is larger at FCC energy in comparison to the LHC energies. In other words, the thermal radiation dominates the direct photon spectra upto a larger $p_T$ value at FCC than at the LHC energies.

The photon elliptic flow parameter $v_2$ as a function of $p_T$ is shown in Fig.~\ref{fig5}. For 0--20\% central collisions the thermal photon  $v_2$ for all three energies are found to be small and close to each other. We see the photon $v_2$ at FCC is slightly smaller for $p_T <$ 2.5 GeV than the elliptic flow calculated at the two LHC energies. 
As we move toward peripheral collisions,  the thermal photon $v_2$ is found to be larger for the higher beam energies. The difference between the $v_2$ values (as a function of $p_T$) at the three energies increases as we go to more peripheral collisions. The prompt photons produced in these collision do not contribute directly to the photon $v_2$, however they dilute the thermal $v_2$ by adding extra weight at the denominator of the photon $v_2$ calculation (see Eq. (9) of Ref.~\cite{chre3}). The direct photon $v_2$ is also plotted in Fig.\ref{fig5} for a comparison.
We see a significant decrease in the elliptic flow for direct photons at all three energies compared to the thermal photon $v_2$.

\section{Summary and conclusions}
We predict the  direct photon transverse momentum spectra from Pb+Pb collisions at 
5.02A TeV at the LHC and at 39A TeV at the proposed Future Circular Collider Facility at CERN. 
The prompt photon production is estimated using a NLO pQCD Monte-Carlo code JETPHOX where the scales of factorization, renormalization, and
fragmentation are set equal to $p_T$  of the photon. We have used CTEQ6.6 parton distribution function, 
BFG-II parton-to-photon fragmentation function and EPS09 parameterization of nuclear shadowing function in this work. 
Thermal photon spectra and elliptic flow  are calculated using a (2+1) dimensional longitudinally boost invariant ideal hydrodynamic 
framework and state-of-the-art photon rates. 
 The prompt photon production is found to be significantly enhanced in Pb+Pb collisions at 39A TeV in comparison to 
5.02A TeV and 2.76A TeV for all three centrality bins. The enhancement factor ranges between 5 to 15 in the $p_T$ region 2 to 15 GeV. 
The time evolutions of average transverse flow velocity and average temperature are also found to be significantly larger at 39A TeV compared to the two LHC energies. 
However, the spatial anisotropy  $\epsilon_x$ as a function of  $\tau$ is found to be smaller for 39A TeV than at 5.02 TeV  and 2.76A TeV.
The momentum anisotropy parameter $\epsilon_p$ is found to be slightly smaller for 39A TeV, otherwise close to each other for all the three energies. 

The thermal photon production is found to be enhanced by a large factor for Pb+Pb collisions at 39A TeV compared to 5.02A TeV and 2.76A TeV. 
However we notice that the relative enhancement in prompt photon production compared to thermal photons is more  in peripheral 
collisions for all beam energies. For example, the prompt to thermal photon ratio is $\sim$ 10 at $p_T =$ 6 GeV for 40--60\% Pb+Pb 
collisions at FCC whereas, the ratio is close to 4 for 0--20\% centrality bin. 
The direct (thermal+prompt) photon transverse momentum spectra for Pb+Pb collisions at 2.76A TeV from our calculation are found to explain the ALICE experimental data well in the region $p_T >$ 2 GeV.
The $v_2$ of thermal photons at FCC energy is found to be slightly larger than the $v_2$ at 
the other two LHC energies in the region $p_T >$ 2 GeV. The direct photon $v_2$  is estimated by adding 
the prompt contribution to the photon yield, and we see elliptic flow of photons decreases significantly in the high $p_T$ region. We have seen that the thermal and prompt photon production and the elliptic flow of the thermal photons change at a differing rates as the energy (and the resulting initial conditions) of the collision increases. Thus a simultaneous description of the direct photon spectra and their elliptic flow will put strong constraints on the theoretical description. This should prove to be quite valuable.
\begin{acknowledgments} 
We thank Computer  and Informatics group, VECC for providing computer facility. PDG is grateful to Department of Atomic
Energy, Government of India for financial support. SD gratefully acknowledges the hospitality provided by 
VECC during his visit. DKS gratefully acknowledges
the grant of Raja Ramanna Fellowship by the Department of Atomic Energy, India and support from the ExtreMe Matter Institute EMMI at the GSI Helmholtzzentrum f\"ur Schwerionenforschung, Darmstadt, Germany.
\end{acknowledgments}

\end{document}